%% file: qself2015.tex
\documentclass[conference]{IEEEtran}
\usepackage{cite}
\usepackage[show]{chato-notes} 

\ifCLASSINFOpdf
  \usepackage[pdftex]{graphicx}
  \graphicspath{{./images/}}
  \DeclareGraphicsExtensions{.pdf,.jpeg,.png}
\else
\fi

\usepackage{url}
\usepackage{subfigure}
\def\etal{{\it et al.}}

\begin{document}
%
\title{360$^{\circ}$ Quantified Self}

\author{
\IEEEauthorblockN{Hamed Haddadi, Ferda Ofli, Yelena Mejova, Ingmar Weber, Jaideep Srivastava}
\IEEEauthorblockA{Qatar Computing Research Institute\\Hamad bin Khalifa University}
Email: {{hhaddadi,fofli,ymejova,iweber,jsrivastava}@qf.org.qa}
}


\maketitle

\begin{abstract}
\input{abstract}
\end{abstract}

%
\IEEEpeerreviewmaketitle

\input{intro}

\input{social_media}
\input{wearables}

\input{challenges}
\input{opportunities}


%
%



\bibliographystyle{IEEEtran}

\bibliography{IEEEabrv,hdi}

\end{document}

%% file: abstract.tex
Wearable devices with a wide range of sensors have contributed to the rise of the \emph{Quantified Self} movement, where individuals log everything ranging from the number of steps they have taken, to their heart rate, to their sleeping patterns. Sensors do not, however, typically sense the social and ambient environment of the users, such as general life style attributes or information about their social network. This means that the users themselves, and the medical practitioners, privy to the wearable sensor data, only have a narrow view of the individual, limited mainly to certain aspects of their physical condition.

In this paper we describe a number of use cases for how social media can be used to complement the check-up data and those from sensors to gain a more holistic view on individuals' health, a perspective we call the \emph{360$^{\circ}$ Quantified Self}. Health-related information can be obtained from sources as diverse as food photo sharing, location check-ins, or profile pictures. Additionally, information from a person's ego network can shed light on the social dimension of wellbeing which is widely acknowledged to be of utmost importance, even though they are currently rarely used for medical diagnosis. We articulate a long-term vision describing the desirable list of technical advances and variety of data to achieve an integrated system encompassing Electronic Health Records (EHR), data from wearable devices, alongside information derived from social media data.

%% file: intro.tex
\section{Introduction}
\label{sec:intro}

Increasing healthcare costs and the alarming rise in societal issues such as obesity, diabetes, and mental health in an ageing population have encouraged a large number of initiatives in promoting a healthier lifestyle. These efforts have been supported by the commercial sector and the Quantified Self movement. With the availability of a range of smartphone applications and dedicated wearable devices, an increasing number of individuals monitor their physical activity levels, dietary intake, vital signs and health status on a continuous basis. The use of Quantified Self applications is soaring also due to individuals' eagerness to share their data (e.g., running times, distance, route, calories intake, etc.) with peers on social media. In addition, data from the individuals' activity on social media, (e.g., sentimental tweets shared, Instagram posts of food pictures, FourSquare check-ins, etc.) can be a valuable signal about individuals' health. 

The research and medical community have also been promoting the Quantified Self movement and self-tracking~\cite{Paton:2012uj, Pearson:2011ct, Swan:2012gn}. Availability of such a large collection of wearable devices and apps can greatly help physicians in understanding the activity levels of patients instead of relying on self-reports. Diabetic patients can perform daily blood glucose level and blood pressure monitoring, and self-report them to their doctors via new standardization efforts, apps and platforms such as Open mHealth\footnote{\url{http://www.openmhealth.org/developers/schemas/}} and accessible smartphone data collection methods such as the Apple HealthKit\footnote{\url{https://developer.apple.com/healthkit/}} and SensingKit~\cite{katevas2014poster}.

Despite their richness, physical activity monitoring devices do not fully capture the health and wellbeing of an individual. They may not be worn at all times, they may fall short in capturing stress levels, diet quality, social interactions, and the effect of one's social network. One may be on an active sports camp, but not be carrying a step counter. Online Social Networks (OSNs), however, present a rich source of information which could complement the physical activity data. OSNs provide rich context into the individuals' social life, from the locations visited, to the foods eaten, and to the emotions experienced. As such, including these data as part of general diagnostics presents a promising opportunity.

Bringing data together from different sources delivers a great promise for the individuals, the health sector, and the society as a whole. Doctors can have a better temporal view of the health trends of the patients, supporting their judgements based on the standard routine check-ups at clinics. Individuals can gain richer insights and more intuitive view into their own wellbeing and discuss the trends with the physician, and the society can benefit from the range of large-scale data available to the research community which can lead to better understanding of the significant factors effective in improving physical and mental health.

\begin{figure}[t]
	\begin{center} 
		\includegraphics[width=1.0\linewidth]{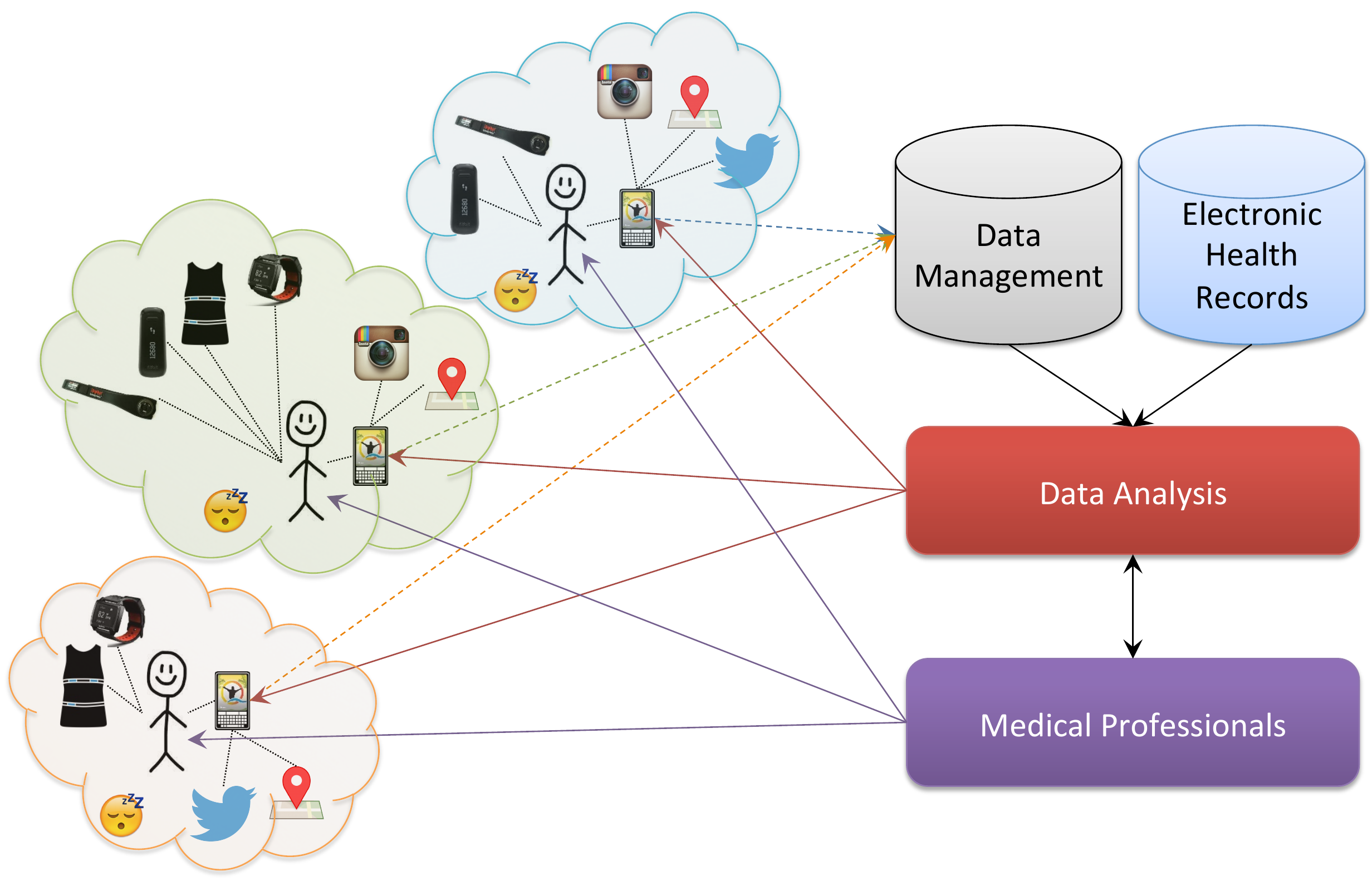}
		\caption{Illustration of the proposed 360$^{\circ}$ Quantified Self view.}
		\label{fig:system_overview}
	\end{center}
\end{figure}

In this paper we offer our vision for a holistic view of the individuals using rich data available from three main sources: OSNs, wearable devices and apps, and Electronic Health Records (EHR). We refer to this view as the \emph{360$^{\circ}$ Quantified Self}, enabling a thorough analysis of each individuals' physical and mental health. Fig.~\ref{fig:system_overview} illustrates the proposed 360$^{\circ}$ Quantified Self view. In the rest of the paper we briefly review the building blocks of this holistic approach, present our ongoing efforts at addressing some of the elements of this vision, and examine the challenges and opportunities they present.

%% file: social_media.tex
\section{Social Media and Health}
\label{sec:social_media}

The earliest uses of social media in the health sector centered around its potential as a communication medium \cite{chou2009social,hawn2009take}. Though this is still an angle worth exploring, in particular when it comes to designing \emph{interventions}, we limit our discussion to the use of social media data for \emph{health monitoring and analysis}.

Most existing works in this space use social media for \emph{public health} monitoring. This approach is generally disease-centric where for a given disease, such as flu, information on social media is identified. Similar in spirit to Google Flu Trends \cite{dugas2012google}, different researchers have looked at the feasibility of using Twitter data to track changes in flu activity over time \cite{lampos2010tracking,aramaki2011twitter,LambPD13,KostkovaSL14}. The basic approach here involves collecting tweets for certain symptom-related terms such as ``flu'', ``headache'' or ``running nose'' and then correlating time series of matching tweet volumes with time series for CDC-reported levels of flu activity. Tracking dengue fever \cite{gomideetal11websci} was done using similar techniques but with a stronger focus on spatial patterns in addition to temporal ones. Looking exclusively at geographical patterns, different researchers have tried to quantify county-level dietary health \cite{Culotta:2014:ECH:2556288.2557139,abbaretal15chi}, and even population-level depression \cite{DeChoudhury:2013:SMM:2464464.2464480}. A recent survey of the use of Twitter data for public health monitoring can be found in \cite{Kostkova15TwitterSocioscope}.

More recently, researchers have explored the use of social media for \emph{individual health} monitoring. Here, the approach is a patient-centric one where for a given patient the goal is to gather data and identify potential symptoms and diseases. Although, unlike the aggregate statistics used for public health monitoring, such individual-level analysis is much more prone to sparsity issues, notable advances in the area have recently been made.

One such area is \emph{mental health} \cite{balanidechoudhury15chi,CoppersmithHD14,DeChoudhury:2013:PPC:2470654.2466447}. As opposed to physical health, no explicit mention of a symptom such as depression is required here -- though still helpful -- and using techniques such as sentiment analysis or even changes in tweeting frequency can give indications concerning underlying changes in mood. Alternatively, popular online forums such as Reddit contain a high amount of self-disclosure which can be automatically detected \cite{balanidechoudhury15chi}.

Users' physical health problems, on the other hand, are usually diagnosed by filtering the data on illness-specific terms, with the basic assumption that users not explicitly referring to a symptom or disease are healthy. This assumption applies in particular to\emph{acute} diseases, but it is less likely that someone regularly tweets about \emph{chronic diseases}, for example, ``my blood pressure is still high''. In our research, we track such ``silent'' diseases -- obesity or diabetes -- which require new ways of inferring a user's health status. 

Beyond visible symptoms, \emph{lifestyle} has widely been linked with chronic diseases, including stroke, diabetes, and even some cancers \cite{willett2006prevention}. Social media allows us to glimpse a user's lifestyle and provide a more holistic view of the day-to-day activities which in long term may impact their health. For example, Sadilek and Kautz~\cite{sadilekkautz13wsdm} use Twitter to quantify the impact of social status, exposure to pollution, and interpersonal interactions on one's general state of health. More specifically, Chunara \etal~\cite{chunaraetal13pone} study obesity using Facebook data, showing sedentary interests to be more popular in areas with higher obesity rates. Abbar \etal~\cite{abbaretal15chi} observe similar patterns using data from Twitter. 

When it comes to \emph{social influence} within the social networks in the context of health, focus often falls on addictions and certain disorders. Smoking is the ``poster child'' social disease, as smokers are often introduced to smoking through friends and peer pressure. However, to the best our knowledge, there have not been any studies using public social network data to study the spread of smoking in social networks. A potential road block here is the inference if someone is a smoker, though image-based techniques could potentially be applied \cite{bienlin12icmv,5597770}. Eating disorders are also known to be linked to social influence, as, for example, in the case of anorexia\cite{DeChoudhury:2015:ATC:2750511.2750515,yomtovetal12jmir}. Christakis and Fowler have further suggested that obesity might be a social disease where one person ``infects'' another if they spend enough time together \cite{christakisfowler07nejm}. Indeed, network assortativity of sharing unhealthy food references on Twitter has been observed by Abbar \etal~\cite{abbaretal15chi}, though their data does not allow for any causal inference.

The network-related use-cases above focus on understanding the disease mechanisms. Yet, we also envision potential uses for diagnostic and treatment purposes. For example, a doctor treating a patient who has been suffering from high blood pressure might want to know if the patient's partner is stressed, or if their dependants are struggling in school. For a doctor supporting someone on a particular diet it might be helpful to know if the patient has contacts who are interested in social and sporting activities.

%% file: wearables.tex
\section{Quantified Self}
\label{sec:wearables}

Wearable devices have been at the heart of the Quantified Self movement with an increasing popularity and adoption in the last few years. Advances in display technology, lightweight yet rigid material, increased computational power, ubiquitous connectivity, and new designs as fashionable items, have all given a boost to this popularity. According to a survey by Vandrico Inc.\footnote{http://vandrico.com/wearables}, there are currently more than 300 wearable devices in the market, available from 50 USD to more than 1000 USD with average price being around 300 USD. Majority of these devices fall under ``lifestyle" and ``fitness" categories while the rest spreads across ``entertainment," ``gaming," ``medical," and ``industrial" categories. Accelerometer being the most popular sensory component, a large number of these devices support features such as step counting, activity recognition, caloric expenditure calculation, heart and breathing rate monitoring, skin conductivity measurement, and sleep quality assessment. Some of these devices can even track users' moods, interests, and the noise and light level of the surrounding environment.

Wearable devices generate a wealth of data for the individuals interested in the Quantified Self movement. In a sense, the importance of health in the individual context has overshadowed other aspects of the Quantified Self movement to an extent that some believe that health is the only objective of this movement. Though for majority of individuals, this trend does not go beyond plots of their daily step counts, and eventually leads to abandoning their wearable device due to lack of more useful feedback. In addition, the dominant inaccessibility of raw data from the majority of these devices has made it difficult for an ecosystem to be developed around them.

Nevertheless, there have been recent attempts in the research community toward using data from unobtrusive sensing and wearable devices for patient monitoring and health informatics systems. Zheng \etal~\cite{Zheng:TBE14ev} provide an overview of these emerging technologies that are essential to the realization of pervasive health monitoring systems. Early examples of such mobile sensing technologies, however, lack one or more of the essential properties such as security, providing high availability, and supporting multiple third-party health-related applications that share access to individuals' devices and data. In their recent paper, Molina-Markham \etal~\cite{MolinaMarkham:MMA14ht} try to address this issue by proposing a secure system architecture (Amulet) for a low-powered bracelet that can run multiple applications and manage access to shared resources in a body-area health network.

\begin{figure*}
  \subfigure[Activity breakdown]{%
    \includegraphics[width=0.32\textwidth]{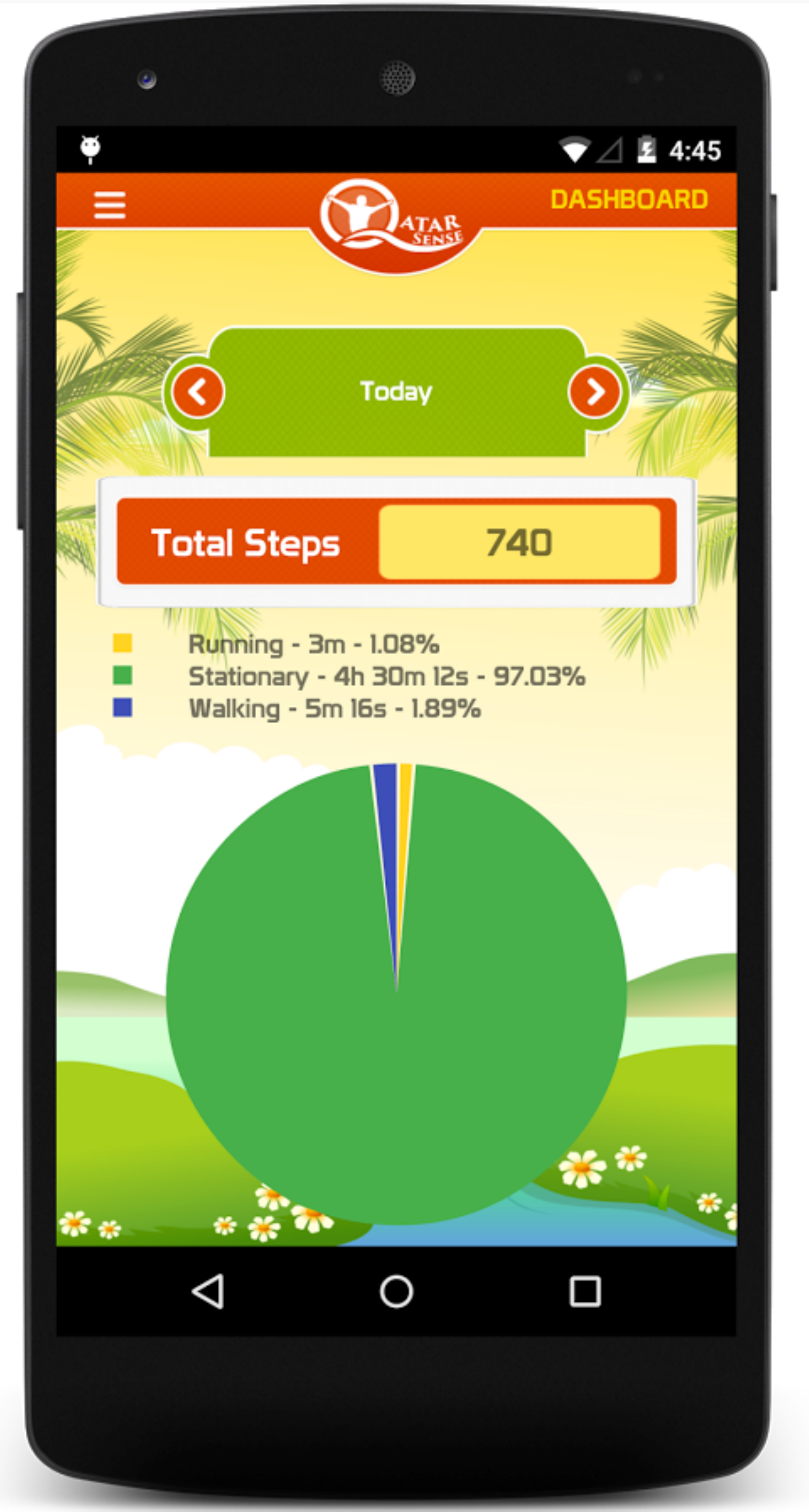}
    \label{map:cnt}
  }
  \hfill
   \subfigure[Leaderboard view]{%
    \includegraphics[width=0.32\textwidth]{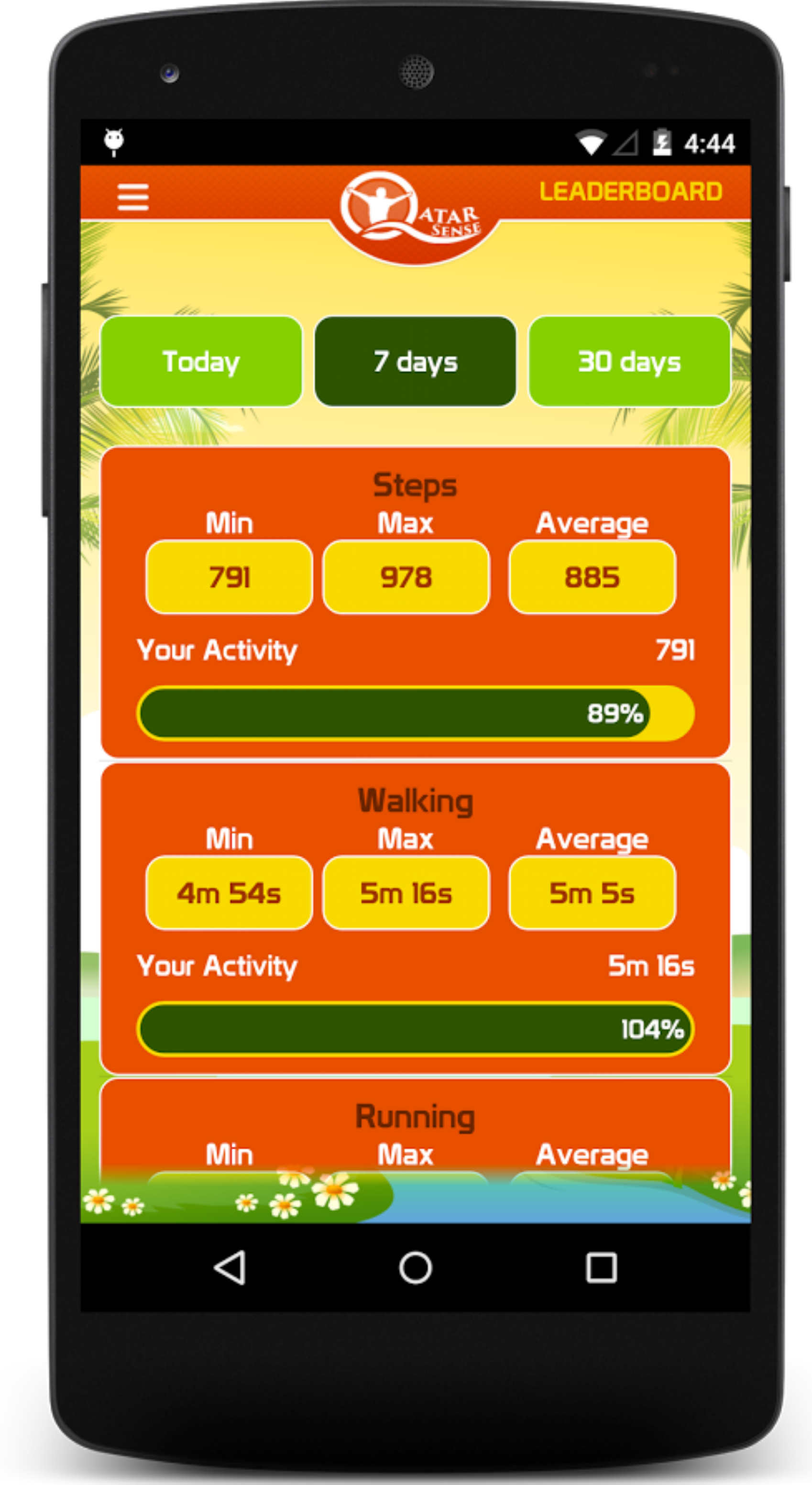}
    \label{map:geo_alexa}
  }
    \subfigure[Notifications and interventions]{%
    \includegraphics[width=0.32\textwidth]{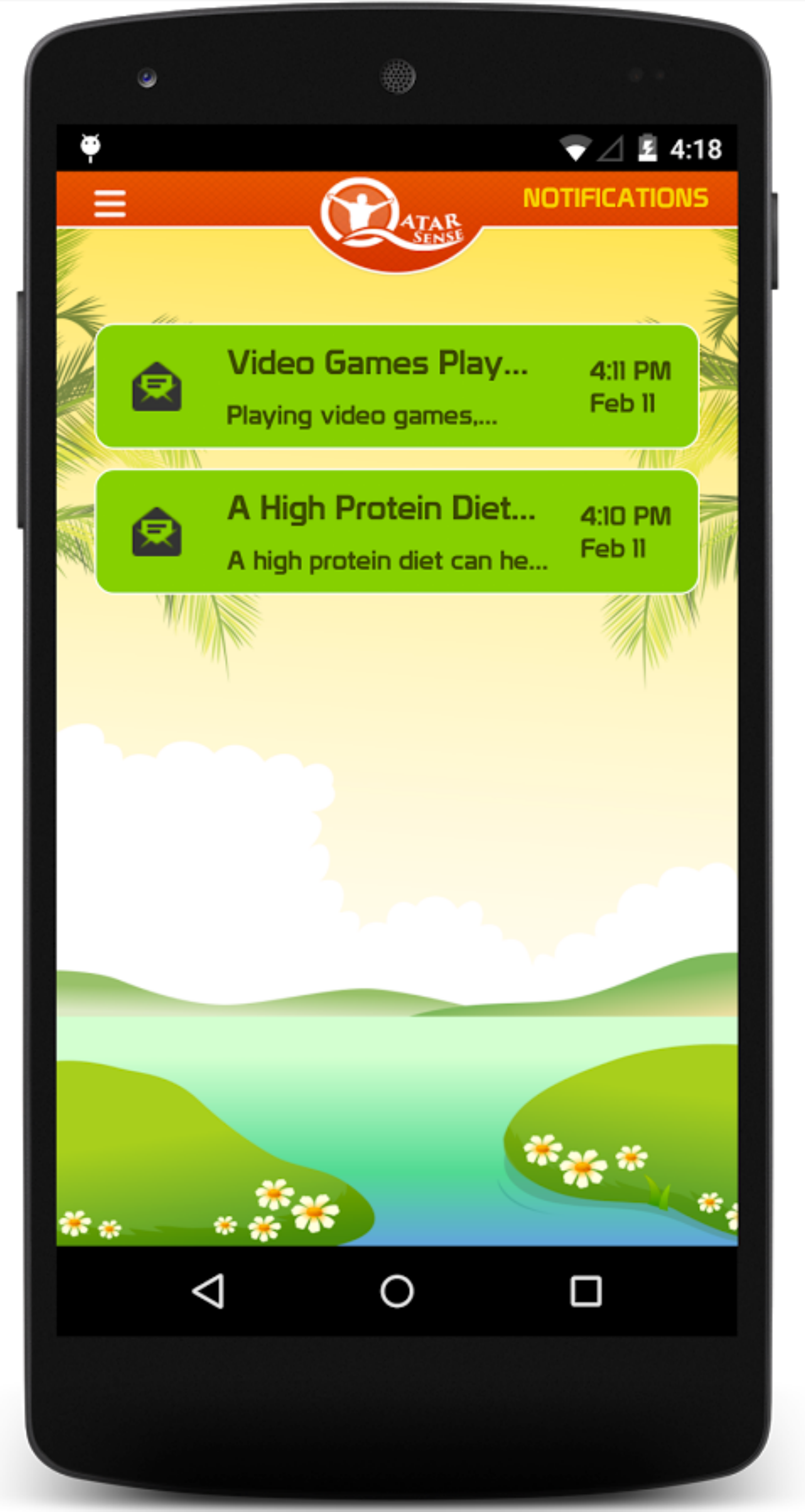}
    \label{map:geo_geo}
  }
  \caption{\label{qatarsense}The QatarSense app interface for feedback and interaction with young children.}
\end{figure*}

There have been a number of attempts on using mobile devices as activity, environment, and emotion sensing context (e.g.,~\cite{Rachuri:2010:EMP:1864349.1864393}). Interventions, encouragements, leaderboards, and intuitive feedback can also help with improving the users' engagement with their apps. In ongoing research into addressing childhood obesity, we are using the QatarSense Android app\footnote{\url{https://play.google.com/store/apps/details?id=com.qcri.qsense&hl=en}} to achieve some of these factors. Figure~\ref{qatarsense} shows the simple interface of this app which also enables social inclusion via a personalised activity leaderboard, and customised interventions and feedback via the doctors for the specific user. In combination with specific ERH details of each user, doctors are able to observe and monitor their activity level, and provide feedback to the children and their parents on potential for improving their health.

In a more clinical setting, Clifton \etal~\cite{Clifton:JBHI14ku} have recently explored combining routine clinical observations with continuous data acquired from mobile pulse oximeters and ECG sensors to provide early warning of physiological deterioration such that a degree of preventive care may be provided to improve patient outcomes. The results of their clinical studies showed that the proposed predictive patient monitoring system outperformed the conventional early warning score (EWS) system by a margin of up to 13\% less false-positive rate.

There are also other recent studies that investigate the mental health in addition to the physical health of the individuals using mobile sensing (mainly smartphone) technologies. For instance, Lane \etal~\cite{Lane:MNA14iu} developed a smartphone application (BeWell+), which monitors users' sleep, physical activity and social interaction to promote improved behavioral patterns via feedback rendered as an ambient display on the smartphone's wallpaper. Similarly, Wang \etal~\cite{Wang:UbiComp14iy} developed a smartphone app (StudentLife) to use automatic and continuous smartphone sensing to assess mental health, academic performance and behavioral trends of the students. Both studies provide important insights into behavioral trends as well as correlations between sensor data from smartphones and mental wellbeing of individuals. A more comprehensive survey on mobile phone sensing can be found in~\cite{lane2010survey}.

Beside the aforementioned research settings, wearable devices and Quantified Self enthusiasts have also seen a popularity in integration with social media, despite the personal nature of the wearable technology and the Quantified Self movement. Many individuals share their running maps and times on Facebook and Twitter directly from apps such as MyFitnessPal. This trend calls for further investigation of the reasons behind such \emph{social health} activity sharing. Do individuals wish to gain support from their social network and friends? Do they wish to hold themselves accountable to their stated health goals? Though the effectiveness of public goal-setting has been debated~\cite{gollwitzer2009intentions}, the inherently social nature of individuals' personal health data and their interaction with it needs to be investigated using established methods in order to understand the long-term effects of such sharing practices~\cite{mortier2014human}.

 
 

%% file: challenges.tex
\section{Challenges}
\label{sec:challenges}

Integrating social media data, physical condition data, and EHRs will face a set of challenges from technical, societal, and legal perspectives~\cite{jaideep}. We briefly overview some of these challenges.

\subsection{Technical Challenges}

Bringing disparate, and often sparse, data from different data sources has a number of technical challenges. People do not always carry their physical activity monitors, nor post about every single mood swing they go through or snap every single item they eat. Hence inferring context, stress level, mood, calorie intake, and physical activity level from these sources requires sophisticated data analysis techniques. For instance, examining food consumption patterns from Instagram and Twitter can shed light on the societal eating habits~\cite{abbaretal15chi, mejova2015foodporn}, however inferring individual calorie intake (e.g., from food images) without consistent manual input is a difficult step. The sparsity issue is more severe for social media data when compared to wearable devices which have a more ubiquitous nature. 


\subsection{Deployment and Inference}

The availability of \emph{Big Data} from all these sources can be a potential burden to already-stressed medical systems in both developing and the developed world, where doctor's time is a scarce resource. At the moment none of the commercially available devices have been regulated by the Food and Drug Administration authority. Making sense from these data sources and correct understanding of their anomalies and inherent biases is also a challenge for the traditional medical domain. Thus, proper analysis and visualization are imperative in transforming the raw sensor data to actionable information medical professionals can use. Furthermore, education in the data gathering and quality would further help the integration of these new sources of information to be integrated into the healthcare domain. The current range of implantable or wearable medical devices also face security challenges from adversaries(see~\cite{Sametinger:2015:SCM:2749359.2667218} for a detailed discussion). These devices are often optimized for functionality and efficiency, rather than security, hence their vulnerabilities can subject them to data manipulation attacks.


\subsection{Ethics and Privacy}

The highly sensitive and private nature of health data pose a number of ethical challenges for ubiquitous monitoring using wearable devices and social media~\cite{brown2007ethical, brown2015social}. Sharing of these data between different providers, and even the medical professionals, introduces a new level of challenges with the increased level of cross-inferences possible across disjoint datasets. A number of solutions, such as use of anonymization techniques~\cite{Loukides27042010} and user-controlled aggregation points such as the Databox~\cite{haddadi2015personal} have been proposed in order to address some of these challenges by providing privacy-preserving methods of accessing and analyzing otherwise scattered pieces of information.

%% file: opportunities.tex
\section{Opportunities}
\label{sec:opportunities}

In this paper we have presented some potential scenarios in which the aggregation of of disparate sources of information, mainly wearable devices, EHRs, and social media content, can improve and potentially transform the current trends in personal and public health and wellness. Availability of such large-scale data form a variety of source, if collected and dealt with responsibility and carefully, presents a great opportunities for unprecedented advances in healthcare and wellness research. We have presented some recent of the recent research in this space and our ongoing efforts in data fusion form different sources in order to improve our understanding of the individuals' overall wellbeing.

One can envision new opportunities in personal health and understanding correlation and causations between physical and mental health (e.g., using data from an individuals' EHR, prescribed medication, and post-hoc sentiment analysis of their social media content), or public health (understanding relationship between mental health or moods, and natural conditions~\cite{hannak2012tweetin} or financial situations). Privacy challenges remain a major obstacle to wide-scale use of personal data for public health inference, though advances in large-scale privacy-preserving analysis techniques such as distributed Differential Privacy~\cite{privalytics} and secure personal data storage facilities can potentially mitigate the privacy issues.

One of the main objectives of the Quantified Self and e-health technologies is the provision of effective behavioural interventions for promoting better health~\cite{info:doi/10.2196/resprot.3990}. Similarly, the more holistic healthcare systems will not solely rely on single-sourced data points such as blood pressure and heart rate. To this end, we believe the aggregation of various form of \textit{Small} (personal) data~\cite{Estrin:2014:SDN:2580723.2580944} under the 360$^{\circ}$ Quantified Self architecture can provide a wealth of additional benefits when compared to each of the data components in isolation.

 
